\documentclass[aps,pra,showpacs]{revtex4}
\begin{document}

\title{Prepotential approach to exact and quasi-exact solvabilities}

\author{Choon-Lin Ho
}
\affiliation{Department of Physics, Tamkang
University, Tamsui 251, Taiwan, Republic of China}


\begin{abstract}

Exact and quasi-exact solvabilities of the one-dimensional
Schr\"odinger equation are discussed from a unified  viewpoint
based on  the prepotential together with Bethe ansatz equations.
This is a constructive approach which gives the potential as well
as the eigenfunctions and eigenvalues simultaneously. The novel
feature of the present work is the realization that both exact and
quasi-exact solvabilities can be solely classified by two
integers, the degrees of two polynomials which determine the
change of variable and the zero-th order prepotential.  Most of
the well-known exactly and quasi-exactly solvable models, and many
new quasi-exactly solvable ones, can be generated by appropriately
choosing the two polynomials. This approach can be easily extended
to the constructions of exactly and quasi-exactly solvable Dirac,
Pauli, and Fokker-Planck equations.

\end{abstract}


\maketitle

\section{Introduction}

Two decades ago, a new class of potentials which are intermediate
to exactly solvable (ES) potentials and non solvable ones have
been found for the Schr\"odinger equation. These are called
quasi-exactly solvable (QES) models for which it is possible to
determine algebraically a part of the spectrum but not the whole
spectrum
\cite{TU,Tur1,Tur,Ush1,Ush,O1,O2,O3,Tur2,Tur3,Tur4,S1,S2}. The
discovery of this class of spectral problems has greatly enlarged
the number of physical systems which we can study analytically. In
the last few years, QES theory has also been extended to the Pauli
\cite{HoRoy1} and Dirac equations
\cite{B1,B2,Znojil,Ho1,Ho2,Ho3,Ho4,Ho5,SH}. More recently, we have
considered QES quasinormal modes, which are damping modes with
complex eigen-energies \cite{CH}. Such modes are of interest in
black hole physics.

Usually a QES problem admits a certain underlying Lie algebraic
symmetry which is responsible for the quasi-exact solutions.  Such
underlying symmetry is most easily studied in the Lie-algebraic
approach \cite{Tur,O1,O2,O3,Tur2,Tur3,Tur4,S1,S2}. However,
solutions of QES states are more directly found in the analytic
approach based on the Bethe ansatz equations
\cite{Ush,Ush1,Gaudin}.  In this analytic approach the form of the
wave functions containing some parameters are assumed from the
very beginning, and these parameters are fitted to make the ansatz
compatible with the potential under consideration . Further
developments in QES theory include classification of
one-dimensional QES operators possessing finite-dimensional
invariant subspace with a basis of monomials \cite{PT}, and
formulation extending to nonlinear operators \cite{KMO}.

A different direction in the development of QES theory is the
prepotential approach \cite{ST}. Here the emphasis was shifted
from the potential to the so-called prepotential (or
superpotential), a concept which plays a fundamental role in
supersymmetric quantum mechanics \cite{Cooper,Junker} (we mention
here that the prepotential $W$ considered here is the integral of
the superpotetial $W$ in \cite{Cooper,Junker,HoRoy1,Ho4,Ho5}).
Prepotential has been extensively employed to study classical and
quantum integrability in Calogero-Moser systems
\cite{CS,Sa1,Sa2,Sa3,Sa4,Sa5}. The simplest prepotentials are
those which give rise to QES potentials admitting just the ground
states.  Physically these are factors which take care of the
asymptotic behaviors of the systems.  They are called the gauge
factors in \cite{Tur1,Tur} (which we shall call the zero-th order
prepotential).   Classification of all possible gauge factors for
$sl(2)$-based QES systems is presented in \cite{O2}. Unlike
previous works, however, in the prepotential approach of \cite{ST}
the prepotential assumes a more fundamental role.  From a
different consideration, it was found that the Schr\"odinger QES
theory was most easily extended to systems with multi-component
wave functions by recasting the Lie-algebraic theory in terms of
the prepotentials \cite{HoRoy1,Ho4,Ho5}. More recently, QES theory
was extended to the Fokker-Planck equations also via the
prepotential approach \cite{HS}.

It thus seems that the prepotential approach to QES theory
deserves a more in-depth study than it has received so far. The
merit of this approach is that the form of the potential of the
system concerned needs not be assumed from the beginning.  All
information about the system is contained in the prepotential and
the solutions, or roots, of the Bethe ansatz equations. The
prepotential and the roots determine the potential as well as the
eigenfunctions and eigenvalues simultaneously. Also, in this
approach exact and quasi-exact solvabilities can be treated on the
same footing. Furthermore, as mentioned in the last paragraph,
such approach facilitates extension of the QES theory from the
Schr\"odinger equation to equations for multi-component wave
functions.

The ideas of the prepotential approach to exact and quasi-exact
solvabilities have been presented in \cite{ST,HoRoy1,Ho4,Ho5}, and
summarized in \cite{HS}.  The emphasis of these works was placed
on the feasibility and elegance of the prepotential approach.
However, the forms of the the prepotential and the required change
of coordinates were either directly adapted from the known ES and
QES models, or given as known for the new QES systems.  Now we
would like to give a first attempt to address the questions as to
how the choice of coordinate transformation and the prepotential
are determined, at least for certain classes of coordinates and
prepotentials.

The main result of this work is the realization that exact and
quasi-exact solvabilities can be solely classified by two
integers, the degrees of the two polynomials which determine the
change of variable and the zero-th order prepotential. Such
classification scheme has not been explored before.  There are
upper limits for theses two integers beyond which no ES systems
are possible.   By selecting appropriate values of these two
degrees, most of the well-known ES and QES models, and many new
QES ones, are easily generated. This approach does not rely on the
knowledge of possible existence of any underlying symmetry in the
system concerned.  It treats both exact and quasi-exact
solvabilities on the same footing.

This paper is organized as follows.  In Sect.~II we present the
idea of the prepotential approach to the exact/quasi-exact
solvability of the Schr\"odnger equation. The general conditions
which the required change of variable and the choice of the
prepotential are discussed.  Two common types of transformations
of coordinates leading to exact/quasi-exact solvable systems are
discussed and examples presented in Sect.~III and IV.  In Sect.~V
and VI the prepotential is generalized to systems defined on the
half-line and on finite interval, respectively. Sect.~VII
concludes the paper.

\section{Prepotential approach}

Suppose $\phi_0(x)~(-\infty<x<\infty))$ is the ground state, with
zero energy, of a Hamiltonian $H_0$: $H_0\phi_0=0$. By the well
known oscillation theorem $\phi_0$ is nodeless, and thus can be
written as $\phi_0\equiv e^{-W_0(x)}$, where $W_0(x)$ is a regular
function of $x$ (this will be assumed in the rest of this paper).
For the square-integrable $\phi_0$, this is the simplest example
of quasi-exact solvability. This implies that the potential $V_0$
is completely determined by $W_0$: $V_0={W_0^\prime}^2 -
W_0^{\prime\prime}$, and consequently, the Hamiltonian is
factorizable (we adopt the unit system in which $\hbar$ and the
mass $m$ of the particle are such that $\hbar=2m=1$):
\begin{equation}
H_0=\left(-\frac{d}{dx}+W_0^\prime\right)\left(\frac{d}{dx}+W_0^\prime\right).
     \label{facHam}
\end{equation}
This fact can be considered as the very base of the factorization
method \cite{fac1,fac2} and of
 supersymmetric quantum mechanics \cite{Cooper,Junker}.  We shall call
$W_0(x)$ the zero-th order prepotential.

Consider now a wave function $\phi_N$ ($N\geq 0$) which is related
to $\phi_0$ of $H_0$ by $\phi_N=\phi_0{\tilde \phi}_N$, where
\begin{eqnarray}
 {\tilde\phi}_N=(z-z_1)(z-z_2)\cdots
(z-z_N),~~{\tilde\phi}_0\equiv 1. \label{phi-2}
\end{eqnarray}
 Here $z=z(x)$ is some function of $x$. In taking the form of
${\tilde\phi}_N$ in Eq.~(\ref{phi-2}) we have assumed that the
only singularities of the system are $z=\pm\infty$.  Other
situations will be addressed later. The function ${\tilde\phi}_N$
is a polynomial in an $(N+1)$-dimensional Hilbert space with the
basis $\langle 1,z,z^2,\ldots,z^N \rangle$.  One can rewrite
$\phi_N$ as
\begin{eqnarray}
 \phi_N =\exp\left(- W_N(x,\{z_k\})
\right), \label{f2}
\end{eqnarray}
with the $N$-th order prepotential $W_N$ being defined by
\begin{eqnarray}
W_N(x,\{z_k\}) = W_0(x) - \sum_{k=1}^N \ln |z(x)-z_k|. \label{W}
\end{eqnarray}
Operating on $\phi_N$ by the operator $-d^2/dx^2$ results in a
Schr\"odinger equation $H_N\phi_N=0$, where
\begin{eqnarray}
H_N &=&-\frac{d^2}{dx^2} + V_N,\\
V_N&\equiv&  W_N^{\prime 2} - W_N^{\prime\prime}.
\end{eqnarray}
From Eq.~(\ref{W}), $V_N$ has the form $V_N=V_0+\Delta V_N$, where
$V_0=W_0^{\prime 2} - W_0^{\prime\prime}$, and
\begin{eqnarray}
\Delta V_N \equiv -2\left(W_0^\prime z^\prime
-\frac{z^{\prime\prime}}{2}\right)\sum_{k=1}^N \frac{1}{z-z_k} +
\sum_{k,l\atop k\neq l} \frac{z^{\prime 2}}{(z-z_k)(z-z_l)}.
\label{dV}
\end{eqnarray}
Here the prime denotes derivative w.r.t. the variable $x$.

$\Delta V_N$ is generally a meromorphic function of $z$ with at
most simple poles. Let us demand that the residues of the simple
poles, $z_{k}$, $k=1,\ldots, N$ should all vanish. This will
result in a set of algebraic equations which the parameters
$\{z_k\}$ must satisfy. These equations are called the Bethe
ansatz equations for $\{z_k\}$.  With $\{z_k\}$ satisfying the
Bethe ansatz equations, $\Delta V_N$ will have no simple poles at
$\{z_k\}$ but it still generally depends on $\{z_k\}$. Thus the
form of $V_N$ is determined by the choice of $z^{\prime 2}$ and
$W_0(x)$.

In what follows we would like to demonstrate that the choice of
$z^{\prime 2}$ and $W_0^\prime z^\prime$ determine the nature of
solvability of the quantal system.   We shall restrict our
consideration only to those cases where $W_0^\prime z^\prime=P_m
(z)$, $z^{\prime 2}=Q_n(z)$ and $z^{\prime\prime}=R_l(z)$ are
polynomials in $z$ of degree $m$, $n$ and $l$, respectively.  Of
course, $Q_n(z)$ and $R_l(z)$ are not independent.  In fact, from
$2z^{\prime\prime}=dz^{\prime 2}/dz$ we have $l=n-1$ and
\begin{eqnarray}
z^{\prime\prime}=R_{n-1}(z)=\frac{1}{2}\frac{dQ_n(z)}{dz}.
\end{eqnarray}
Consequently, the variables $x$ and $z$ are related by (we assume
$z(x)$ is invertible for practical purposes)
\begin{eqnarray}
x(z)=\pm \int^z \frac{dz}{\sqrt{Q_n(z)}},\label{z(x)}
\end{eqnarray}
and the prepotential $W_0(x)$ is determined as
\begin{eqnarray}
W_0(x)=\int^x dx
\left(\frac{P_m(z)}{\sqrt{Q_n(z)}}\right)_{z=z(x)}.\label{W0}
\end{eqnarray}
Eqs.~(\ref{z(x)}) and (\ref{W0}) define the transformation $z(x)$
and the corresponding prepotential $W_0(x)$. Thus, $P_m(z)$ and
$Q_n(z)$ determine the quantum system.  Of course, the choice of
$P_m$ and $Q_n$ must be such that $W_0$ derived from (\ref{W0})
must ensure normalizability of $\phi_0=\exp(-W_0)$.

Now depending on the degrees of the polynomials $P_m$ and $Q_n$,
we have the following situations:

\begin{enumerate}
\item[(i)] if $\max\{m,n-1\}\leq 1$, then in $V_N(x)$
the parameter $N$ and the roots $z_k$'s will only appear as an
additive constant and not in any term involving powers of $z$.
Such system is then exactly solvable;

\item[(ii)] if $\max\{m,n-1\}=2$, then $N$ will appear in
the first power term in $z$, but $z_k$'s only in an additive term.
This system then belongs to the so-called type 1 QES system
defined in \cite{Tur}, i.e., for each $N\geq 0$, $V_N$ admits
$N+1$ solvable states with the eigenvalues being given by the
$N+1$ sets of roots $z_k$'s.  This is the main type of QES systems
considered in the literature;

\item[(iii)] if $\min\{m,n-1\}\geq 3$, then not only $N$ but also
$z_k$'s will appear in terms involving powers of $z$. This means
that for each $N\geq 0$, there are $N+1$ different potentials
$V_N$, differing in several parameters in terms involving powers
of $z$, have the same eigenvalue (when the additive constant, or
the zero point, is appropriately adjusted). When $z_k$'s appear
only in the first power term in $z$, such systems are called type
2 QES systems in \cite{Tur}.  We see that QES models of higher
types are possible.
\end{enumerate}

We will illustrate these general situations with specific examples
in the following sections.  For definiteness, in this paper we
will only consider cases with $z^{\prime 2}=Q_2(z)\equiv
q_2z^2+q_1z+q_0$ (i.e., $z^{\prime\prime}=q_2z+q_1/2$), where
$q_2,~q_1$ and $q_0$ are real constants (by taking $n=2$ here we
mean to include $Q_1$ and $Q_0$ as special cases if some of the
coefficients vanish). This choice of $z^{\prime 2}$ covers most of
the known ES shape-invariant potentials in \cite{Cooper,Junker}
and the $sl(2)$-based QES systems in \cite{Tur}, and a new one
discussed in \cite{ST,HS}. Such coordinates are called
``sinusoidal coordinates", which include quadratic polynomials,
trigonometric, hyperbolic, and exponential types. The connection
of sinusoidal coordinates with ES theory has been extensively
discussed in \cite{OS}.

With this choice of $z^{\prime 2}$, we have
\begin{eqnarray}
 V_N ={W_0^\prime}^2 - W_0^{\prime\prime}+ q_2N^2 -2\sum_{k=1}^N
\frac{1}{z-z_k}\left\{P_m(z) -\frac{q_2}{2}z_k- \frac{q_1}{4} -
\sum_{l\neq k} \frac{Q_2(z_k)}{z_k-z_l}\right\}. \label{dV1}
\end{eqnarray}
In deriving Eq.~(\ref{dV1}) use has been made of the following
identities:
\begin{eqnarray}
\sum_{k,l=1\atop k\neq l}^N \frac{1}{(z-z_k)(z-z_l)} &=& 2
\sum_{k,l=1\atop k\neq l}^N
\frac{1}{z-z_k}\left(\frac{1}{z_k-z_l}\right),\\
\sum_{k,l=1\atop k\neq l}^N \frac{z}{(z-z_k)(z-z_l)} &=& 2
\sum_{k,l=1\atop k\neq l}^N
\frac{1}{z-z_k}\left(\frac{z_k}{z_k-z_l}\right),\\
\sum_{k,l=1\atop k\neq l}^N \frac{z^2}{(z-z_k)(z-z_l)}
  &=& 2 \sum_{k,l=1\atop k\neq l}^N
\frac{1}{z-z_k}\left(\frac{z_k^2}{z_k-z_l}\right)+ N(N-1).
\end{eqnarray}
 Demanding the residues at $z_k$'s vanish gives the set
of Bethe ansatz equations
\begin{eqnarray}
P_m(z_k) -\frac{q_2}{2}z_k- \frac{q_1}{4} - \sum_{l\neq k}
\frac{Q_2(z_k)}{z_k-z_l}=0,~~k=1,2,\ldots,N. \label{BAE}
\end{eqnarray}
Putting back the set of roots $z_k$ into Eq.~(\ref{dV1}), we
obtain a potential $V_N(x)$ without simple poles. The degree of
$P_m(z)$ determines the nature of the solvability of the system,
namely, for $m=1,2,3,\ldots$, the system is, respectively, ES,
type 1 QES, type 2 QES, and higher types QES, as discussed
generally in the last section.

To proceed further, we must specify $P_m(z)$.  We shall discuss
cases where $z^{\prime 2}$ is linear and quadratic in $z$
separately.

\section{Examples with $z^{\prime 2}(x)=Q_1(z)$}

Let us first consider the case where $z^{\prime 2}(x)=Q_1(z)$,
i.e., $q_2=0$. For definiteness we take $Q_1(z)=4Az +q_0$
($q_1=4A$), where $A,~q_0$ are real constants. This implies
 $z^{\prime\prime}(x)=R_0(z)=2A$.  Hence, the general solution of $z(x)$
is a quadratic form of $x$:
\begin{eqnarray}
z(x)=Ax^2 + Bx + C,~~A,~B,~C {\rm : real\  constants}.
\label{z}
\end{eqnarray}
It is easily checked that $q_0$ is related to $A,~B$ and $C$
through $z^{\prime 2}=4Az +B^2-4AC$.

We now illustrate how some ES and QES models can be constructed in
the prepotential approach by taking different values of $m$.

\subsection{Exactly solvable cases: $m=1$}

Suppose $m=1$ so that $P_1(z)=A_1z + A_0$ ($A_1,~A_0$ real).  By
writing $P_1=A_1(z-z_k)+A_1z_k+A_0$, one obtains from
Eq.~(\ref{dV1}) that
\begin{eqnarray}
\Delta V_N &=& -2A_1\sum_{k=1}^N 1 -
2\sum_{k=1}^N\frac{1}{z-z_k}\left\{P_1(z_k) -A -
\sum_{l\neq k} \frac{Q_1(z_k)}{z_k-z_l}\right\}\nonumber\\
&=&-2A_1N. \label{dV-case1}
\end{eqnarray}
The last term in braces in Eq.~(\ref{dV-case1}) vanishes when
$z_k$'s satisfy the Bethe ansatz equations (\ref{BAE}).   Now $N$
only appears as a parameter in an additive term, and not in terms
involving powers of $z$ in $\Delta V_N$. The roots $z_k$'s do not
appear at all. The additive term can be treated as the eigenvalue.
The Schr\"odinger equation reads
\begin{eqnarray}
\left(-\frac{d^2}{dx^2} + W_0^{\prime
2}-W_0^{\prime\prime}\right)e^{-W_N}=2A_1Ne^{-W_N}.
\end{eqnarray}
We see that the potential $W_0^{\prime 2}-W_0^{\prime\prime}$ is
ES: by varying $N$, one obtains all the eigenvalues $2A_1N$ and
the eigenfunctions $\phi_N=\exp(-W_N)$.

As an example, let us take $Q_1(z)=1$ and $P_1(z)=bz$. A
particular solution of $z$ is $z(x)=x$. From Eq.~(\ref{W0}) one
gets
\begin{eqnarray}
W_0(x)=\int^x \frac{bx}{1}dx=\frac{bx^2}{2}+{\rm const.}
\end{eqnarray}
The integration constant is to be determined by normalization of
the wave function.  For $\phi_0=\exp(-W_0)$ to be
square-integrable, one must assume $b>0$. The BAE (\ref{BAE}) are:
\begin{eqnarray}
bx_k-\sum_{l\neq k} \frac{1}{x_k-x_l}=0, ~~k=1,\ldots,N,
\label{BAE-Osc}
\end{eqnarray}
and the potential is
\begin{eqnarray}
V_0=b^2 x^2 -b,~~\Delta V_N=-2Nb.
\end{eqnarray}
This leads to the Schr\"odinger equation:
\begin{eqnarray}
\left(-\frac{d^2}{dx^2} + b^2 x^2 \right)e^{-W_N}=b(2N+1)e^{-W_N}.
\end{eqnarray}
  This system is just the well-known simple
harmonic oscillator.

We note here that by rescaling $\sqrt{b}x_k\to x_k$,
Eq.~(\ref{BAE-Osc}) will have $b=1$.   The resulted equations are
the equations that determine the zeros of the Hermite polynomials
$H_N(x)$ as found by Stieltjes \cite{Stiel,Szego,CS}. Hence we
have reproduced the well known wave functions for the harmonic
oscillator, namely, $\phi_N=\exp(-W_N)\sim
\exp(-bx^2/2)H_N(\sqrt{b}x)$.

\subsection{Type 1 quasi-exactly solvable cases: $m=2$}

Next we consider $P_2(z)=A_2 z^2 + A_1 z+ A_0$.  By a similar
argument we obtain
\begin{eqnarray}
\Delta V_N= -2A_2 N z -2A_2\sum_{k=1}^N z_k -2A_1 N.
\end{eqnarray}
The Schr\"odinger can be written as
\begin{eqnarray}
\left(-\frac{d^2}{dx^2} + W_0^{\prime 2}-W_0^{\prime\prime}-
2A_2Nz \right)e^{-W_N}=2\left(A_2\sum_{k=1}^N z_k+
A_1N\right)e^{-W_N}. \label{SE-case2}
\end{eqnarray}
Unlike the previous case, now $N$ not only appears in an additive
constant term but also in the term with $z$, and the set of roots
$z_k$'s appear in the additive term. This system is the so-called
type 1 QES models.  Type 1 QES models classified as class VI in
\cite{Tur} belong to this category.

A well-known example is the sextic oscillator, the simplest QES
model of this type \cite{TU}. In our prepotential approach, this
system is defined by $z(x)=x^2$ and $P_2(z)=2(az^2 + bz)$.
 Then
$Q_1(z)=4z$, and
\begin{eqnarray}
W(x)=\int^x \frac{ax^4 + bx^2}{\sqrt{x^2}}dx=\frac{1}{4}ax^4+
\frac{1}{2}bx^2 + {\rm const.}
\end{eqnarray}
Here $a>0$ to ensure square-integrability of the wave function.
The BAE are:
\begin{eqnarray}
2az_k^2 +2bz_k -1 - 4\sum_{l\neq k}\frac{z_k}{z_k-z_l} =0,~~~
k=1,\ldots,N,
\end{eqnarray}
and the potential is
\begin{eqnarray}
V_N=a^2 x^6 +2ab x^4+ \left[b^2-\left(4N +3\right)a\right]x^2
-4a\sum_k z_k -(4N+1)b.
\end{eqnarray}
It is seen that the QES sextic oscillator can be so easily
constructed in the prepotential approach.

\subsection{Type 2 quasi-exactly solvable cases: $m=3$}

We now consider cases with $m=3$ with  $P_3(z)=A_3 z^3 + A_2 z^2 +
A_1 z+ A_0$. Eq.~(\ref{dV1}) leads to
\begin{eqnarray}
\Delta V_N= -2A_3 N z^2 - 2 \left(A_3\sum_{k=1}^N z_k+
A_2N\right)z-2A_3\sum_{k=1}^N z_k^2 -4A_2\sum_{k=1}^N z_k  -2A_1
N.
\end{eqnarray}
Now $N$ appears in $z$ and $z^2$ terms, and also in an additive
constant term. The roots $z_k$'s now not only appear in the
additive term but also in the term with $z$. This is a type 2 QES
model.

A simple example of this type is given by the defining relations
$z(x)=x$ and $P_3(z)=az^3+bz$. The prepotential is
$W_0=ax^4/4+bx^2/2$, which is exactly the same as that for the
sextic oscillator discussed in the last section.  The two models
differ only in the choice of $z(x)$.  The corresponding $V_N$ is
\begin{eqnarray}
V_N=a^2 x^6 +2ab x^4+ \left[b^2-\left(2N +3\right)a\right]x^2 - 2
a\left(\sum_{k=1}^N x_k \right)x-2a\sum_{k=1}^N x_k^2 -(2N+1)b.
\end{eqnarray}
This is a new QES model.

It is now easy to see that, for $m\geq 4$, not only will $N$
appear in more terms involving powers of $z$, but also the set of
roots $z_k$'s. This will give rise to new  general types of QES
systems as mentioned in Sect.~II.

\section{Examples with $z^{\prime 2}(x)=Q_2(z)$}

We now come to cases where $z^{\prime 2}(x)=Q_2(z)$ with $q_2\neq
0$. Here $z(x)$ is again some sinusoidal coordinates, which
include the exponential, hyperbolic and trigonometric functions.
Construction of models proceeds as before. By taking appropriate
$P_m(z)$, one can reconstruct class I, II and X QES models in
\cite{Tur}, and some of the ES models listed in \cite{Cooper},
namely, the Morse potential, the Scarf I and II potentials, and
the P\"oschl-Teller potential. We will not bore the reader by
going through all the cases here. Instead we shall briefly discuss
the Morse potential, as we would like to show how easy its
potential, eigenvalues and eigenfunctions are constructed, and to
compare this construction with another construction based on a
different choice of the prepotential to be discussed in the next
section.

Suppose we take $Q_2(z)=\alpha^2 z^2$, and choose a solution
$z(x)=\exp(\alpha x)$ (henceforth $\alpha$ is taken as a positive
real constant).  Let $P_1(z)=\alpha (Az-B)$ ($A,~B>0$ real
constants). The parametrization is chosen such that the form of
the Morse potential given in \cite{Cooper} is recovered. From
Eq.~(\ref{W0}) we have $W_0^\prime=A-B\exp(-\alpha x)$.   Hence
\begin{equation}
V_N=A^2-B\left(2A+\alpha\right)e^{-\alpha x}+B^2 e^{-2\alpha x}
-\left[A^2-\left(A-N\alpha\right)^2\right],\label{Morse}
\end{equation}
with $z_k$'s satisfying the BAE
\begin{eqnarray}
Az_k -B -\frac{\alpha}{2}z_k-\alpha \sum_{l\neq
k}\frac{z_k^2}{z_k-z_l}=0.\label{BAE-Morse1}
\end{eqnarray}
This is the ES shape-invariant Morse potential listed in
\cite{Cooper}.  Taking the first three terms in Eq.~(\ref{Morse})
as the traditional Morse potential, the eigenvalues are given by
$A^2-(A-N\alpha)^2$, in agreement with the result given
in\cite{Cooper}.  The wave functions are
\begin{equation}
\phi_N(x)\sim \exp\left(-Ax-\frac{B}{\alpha}e^{-\alpha
x}\right)\prod_{k=1}^N(z-z_k), ~~z=e^{\alpha x}.\label{Morse-wf-1}
\end{equation}

Let us make an interesting observation here. From Eq.~(\ref{dV1})
it is obvious that only the coefficients $q_2$ in $Q_2(z)$ and
$p_1$ in $P_1(z)$ will enter the expression of the eigenvalues,
namely, $N(2p_1-q_2N)$.  Now for the Scarf II and the
P\"oschl-Teller potential (both with $Q_2(z)=\alpha^2(1+z^2)$),
$q_2=\alpha^2$ is the same as that in the Morse potential, whereas
for the Scarf I potential (with $Q_2(z)=\alpha^2(1-z^2)$) there is
a sign difference.  So if we choose $P_1(z)$ with the same $p_1$
for all these cases, e.g. $P_1(z)=\alpha A z + p_0$, then we would
expect that the Morse, Scarf II and P\"oschl-Teller potentials
would have the same set of eigenvalues ($A^2-(A-N\alpha)^2$),
while the Scarf I potential has a different set of eigenvalues
differing by a sign in some parameter ($(A+N\alpha)^2-A^2$).  This
is in fact the case \cite{Cooper}. The prepotential approach
presented here gives a very simple and direct explanation of why
this is so.

\section{Prepotentials for systems with singularity $z=0$}

Now we would like to discuss a possible generalization of the
$N$-th order prepotential in Eq.~(\ref{W}) for quantum systems
defined on a half-line (e.g. $x\in (0,\infty)$) with singularity
at the origin.  For such systems, the wave functions may acquire a
prefactor $x^p$, where $p$ is usually some non-negative positive
number, in order to account for the asymptotic behavior at the
origin.  This observation motivates a possible generalization of
the zero-th order prepotential $W_0(x)$ to ${\widetilde
W}_0(x)\equiv W_0(x)-p\ln |z|$ for the ground state
$\phi_0(x)=\exp(-{\widetilde W}_0(x))$. Here $W_0$ is a regular
function of $x$ as before. Eq.~(\ref{W}) becomes
\begin{eqnarray}
W_N(x,\{z_k\}) = W_0(x)-p\ln |z| - \sum_{k=1}^N \ln
|z(x)-z_k|,~~z_k\neq 0. \label{Wp}
\end{eqnarray}
For the moment $p$ is a free parameter.

With the prepotential (\ref{Wp}), the potential
 $V_N=W^{\prime 2}_N-W^{\prime\prime}_N$ has the form
 $V_0+\Delta V_N$ where $V_0$ and $\Delta V_N$ are given by
\begin{eqnarray}
V_0&=&{\widetilde W}^{\prime 2}_0-{\widetilde
W}^{\prime\prime}_0\nonumber\\
&=& W_0^{\prime 2}-W_0^{\prime\prime} -2\left(W_0^\prime z^\prime
-\frac{z^{\prime\prime}}{2}\right)\frac{p}{z}+
p\left(p-1\right)\left(\frac{z^\prime}{z}\right)^2\label{V0p}\\
 \Delta V_N &=& -2\left(W_0^\prime z^\prime
-\frac{z^{\prime\prime}}{2}\right)\sum_{k=1}^N \frac{1}{z-z_k} +
z^{\prime 2}\left(\sum_{k,l\atop k\neq l}
\frac{1}{(z-z_k)(z-z_l)}+\sum_{k=1}^N \frac{2p}{z(z-z_k)}\right) .
\label{dVp}
\end{eqnarray}
Again, the system is completely defined by the choice of
$z^{\prime 2}$ and $W_0$.  However, the presence of terms with the
parameter $p$ changes qualitatively the sufficient conditions
discussed in Sect.~II and opens up new possibilities.

As in previous sections, for definiteness, we shall confine our
discussions here to $z^{\prime 2}=Q_2(z)\equiv q_2z^2+q_1z+q_0$
and $W_0^\prime z^\prime=P_m(z)$.  With these choices, the
potential $V_N$ is given by
\begin{eqnarray}
V_N &=& W_0^{\prime 2}-W_0^{\prime\prime} -p
\left(2P_m(z)-q_1\left(p-\frac{1}{2}\right)\right)\frac{1}{z} +
p\left(p-1\right)q_0\frac{1}{z^2} + q_2\left(N+p\right)^2
\nonumber\\
&& -2 \sum_{k=1}^N \frac{1}{z-z_k}\left\{P_m(z)+\frac{pq_0}{z}
-\left(p+\frac{1}{2}\right)q_2z_k -\left(p+\frac{1}{4}\right)q_1
-\sum_{l\neq k}\frac{Q_2(z_k)}{z_k - z_l}\right\}. \label{dVp-1}
\end{eqnarray}
The corresponding Bethe ansatz equations are
\begin{eqnarray}
P_m(z_k) +\frac{pq_0}{z_k} -\left(p+\frac{1}{2}\right)q_2z_k
-\left(p+\frac{1}{4}\right)q_1 -\sum_{l\neq k}\frac{Q_2(z_k)}{z_k
- z_l}=0.\label{BAE-p}
\end{eqnarray}
It is clear that Eqs.~(\ref{dVp-1}) and (\ref{BAE-p}) reduce to
(\ref{dV1}) and (\ref{BAE}) when $p=0$.

If $q_0=0$ (i.e., $z=0$ is a zero of $z^{\prime 2}$) and $m\leq 1$
one may obtain an ES model. For example, if we take $Q_1(z)=\alpha
z$ and $P_1(z)=Az$, then the three-dimensional oscillator listed
in \cite{Cooper} is recovered. But when $q_0\neq 0$, the presence
of the $pq_0/z(z-z_k)$ term in Eq.~(\ref{dVp-1}) will give rise to
a term $(\sum_k z_k)/z$ in $V_N$, and hence the potential
(\ref{dVp-1}) represents a QES system even if $m\leq 1$, in
contrast to the cases discussed in previous sections. For
instance, if we take $z^{\prime 2}=1$ and $P_m(z)=P_1(z)=a+bz$,
this will produce an ES model if $p=0$, which is the shifted
oscillator $V_N=(bx+a)^2 -(2N+1)b$ (note that the domain of $x$
changes from the half-line to the full line). But for general $p$,
the system is a type 2 QES system classified as the class VIII
system in \cite{Tur}.

However, it is possible that if $p$ assumes certain value, the
nature of the system could be qualitatively changed, such as the
domain of the variable may change from the half-line to the full
line (as in the case of the shifted oscillator mention in the
previous paragraph), or a QES system becomes an ES one. We shall
illustrate these situations with two examples below.

\subsection{Sextic oscillator again}

Following Sect.~III(B), we take $Q_2(z)=4z$ and
$P_m(z)=2(az^2+bz)$. These lead to $z=x^2$ and $W_0(x)=ax^4/4
+bx^2/2 +c$, with real constants $a>0,~b$ and $c$.  The BAE and
$V_N$ are
\begin{eqnarray}
2az_k^2 +2bz_k -\left(4p+1\right) - 4\sum_{l\neq
k}\frac{z_k}{z_k-z_l} =0,~~ k=1,\ldots,N,
\end{eqnarray}
and
\begin{eqnarray}
V_N&=&a^2 x^6 +2ab x^4+ \left[b^2-\left(4N +4p
+3\right)a\right]x^2\nonumber\\
&& +4p\left(p-\frac{1}{2}\right)\frac{1}{x^2} -4a\sum_k z_k
-\left(4N+4p+1\right)b.
\end{eqnarray}
In general, this model is a QES system on the half-line. But if
$p=0$ or $p=1/2$ the  $1/x^2$ term will be absent, and the system
is just the sextic oscillator on the full line. The domain is
extended. The case $p=0$ with symmetric wave functions was
discussed before, which is the class VII QES model in \cite{Tur},
and the case $p=1/2$ with anti-symmetric wave functions was
discussed in \cite{TU,Ush,Ush1}.

\subsection{Morse potential again}

Now we consider the situation which will include the ES Morse
potential as a special case.

We take $Q_2(z)=\alpha^2 z^2$ and $P_m(z)=P_2(z)=\alpha^2z^2/2
-\alpha Az$.  As a special case, we choose the solution
$z=\exp(-\alpha x)$ and $W_0^\prime (x)=A-Bz$ with $B\equiv
\alpha/2$.  As in Sect.~IV, the parametrization is chosen such
that when $p$ assumes special value, the system becomes the Morse
potential given in \cite{Cooper}. The potential is
\begin{eqnarray}
V_N &=& A^2 -B\left(2A + \alpha\right)e^{-\alpha x} + B^2
e^{-2\alpha x} - 2p\alpha\left(\frac{\alpha}{2}z-A\right)
+\alpha^2\left(N+p\right)^2\nonumber\\
&&-2\sum_{k=1}^N \frac{1}{z-z_k}\left\{\frac{\alpha^2}{2}z^2
-\alpha Az -\alpha^2 \left(p+\frac{1}{2}\right)z_k - \sum_{l\neq
k} \frac{\alpha^2 z_k^2}{z_k-z_l}\right\}.
\end{eqnarray}
If we choose $z_k$'s to satisfy the BAE (\ref{BAE-p})
\begin{equation}
\frac{\alpha}{2}z_k - A -\alpha \left(p+\frac{1}{2}\right) -
\alpha\sum_{l\neq k} \frac{
z_k}{z_k-z_l}=0,~~k=1,2,\ldots,N,\label{BAE-Morse2}
\end{equation}
then we arrive at the potential
\begin{eqnarray}
V_N&=&A^2-B\left(2A+\alpha\right)e^{-\alpha x}+B^2 e^{-2\alpha x}
-\alpha^2\left(N+p\right)z\nonumber\\
&&-\alpha^2\sum_k z_k +2\alpha
A\left(N+p\right)+\alpha^2\left(N+p\right)^2.
\end{eqnarray}
The term $-\alpha^2\sum_k z_k$ can be simplified using
Eq.~(\ref{BAE-Morse2}) as
\begin{equation}
-\alpha^2\sum_k
z_k=-\alpha\left[2A+\left(2p+1\right)\alpha\right]N-\alpha^2N\left(N-1\right).
\end{equation}
In general this potential defines a type 1 QES system, as $N$
appears in the first power term of $z$.  This system is not listed
in \cite{Tur}.  But if $p=0$, then the domain of $x$ changes from
the half-line to the full line, and the system becomes that
classified as class I in \cite{Tur}.

It is also obvious that if $p=-N$,  $V_N$ becomes
Eq.~(\ref{Morse}), and the system is the ES Morse potential. Let
us recast Eq.~(\ref{BAE-Morse2}) into the following form (taking
$p=-N$):
\begin{equation}
\sum_{l\neq k} \frac{1}{z_k-z_l} +
\frac{\gamma/2}{z_k}=\frac{1}{2},~~k=1,2,\ldots,N,\label{Laguerre}
\end{equation}
where
\begin{equation}
\gamma\equiv 2\frac{A}{\alpha}-2N+1.
\end{equation}
Eq.~(\ref{Laguerre}) is just the set of equations that determines
the zeros of the associated Laguerre polynomials $L^{\gamma-1}_N
(z)$, i.e., $L^{\gamma-1}_N (z_k)=0$ \cite{Szego,CS}.  Hence the
eigenfunctions $\phi_N=\exp(-W_N)$ are
\begin{eqnarray}
\phi_N(x) &=&\exp\left(-Ax-\frac{B}{\alpha}e^{-\alpha x}\right)
z^{-N}\prod_{k=1}^N(z-z_k)\nonumber\\
 &=&z^{s-N}e^{-\frac{z}{2}}L^{2(s-N)}_N (z),~~s\equiv
\frac{A}{\alpha},~~z=e^{-\alpha x}, \label{Morse-wf-2}
\end{eqnarray}
as given in \cite{Cooper}.

Compared with the discussion in Sect.~IV, it is interesting to see
that the Morse potential can be constructed with two different
prepotentials.  Now one may wonder if the results are consistent,
as the wave functions and the BAE look rather different in these
two constructions.  Below we would like to show that they are
indeed the same.

Let us rewrite Eq.~(\ref{Morse-wf-1}) as
\begin{equation}
\phi_N(x)\sim \exp\left(-Ax-\frac{B}{\alpha}e^{-\alpha
x}\right)z^{N}\prod_{k=1}^N(z^{-1}-z_k^{-1}), ~~z=e^{\alpha
x}.\label{Morse-wf-3}
\end{equation}
We note that in Sect.~IV the variable $z=\exp(\alpha x)$ is
reciprocal to the variable $z=\exp(-\alpha x)$ in this subsection.
Hence when one makes the change $z\to 1/z$ and $z_k\to 1/z_k$ in
Eq.~(\ref{Morse-wf-3}), one arrives at Eq.~(\ref{Morse-wf-2}). Now
one needs only to show that the same transformation in $z$ and
$z_k$ maps the BAE (\ref{BAE-Morse1}) to Eq.~(\ref{BAE-Morse2}).

Making the change $z_k\to 1/z_k$ in Eq.~(\ref{BAE-Morse1}) leads
to
\begin{equation}
A -\frac{\alpha}{2}- Bz_k-\alpha \sum_{l\neq
k}\frac{z_l}{z_l-z_k}=0.\label{B}
\end{equation}
Writing $z_l=(z_l-z_k)+z_k$ in the numerator of the last term in
Eq.~(\ref{B}) and recalling that $B=\alpha/2$, we arrive at
\begin{equation}
\frac{\alpha}{2}z_k - A -\alpha \left(-N+\frac{1}{2}\right) -
\alpha\sum_{l\neq k} \frac{ z_k}{z_k-z_l}=0,~~k=1,2,\ldots,N.
\end{equation}
This is simply Eq.~(\ref{BAE-Morse2}) with $p=-N$.  Thus we have
shown that the wave functions and the BAE obtained in the two
constructions are the same.

\section{Systems defined on finite intervals}

Finally, we consider systems defined on a finite interval. Suppose
the potential of a system is singular at $z=a_1$ and $a_2$, where
$a_1$ and $a_2$ are two real parameters (assuming $a_2>a_1$).
Generalizing the discussions in the last section, it is plausible
to assume for such system a prepotential of the form
\begin{eqnarray}
W_N(x,\{z_k\}) = W_0(x)-p_1\ln |z-a_1|-p_2\ln |z-a_2| -
\sum_{k=1}^N \ln |z(x)-z_k|,~~z_k\neq a_1,~a_2, \label{W-2s}
\end{eqnarray}
where $p_1$ and $p_2$ are two real positive parameters.  The wave
function has the form
\begin{equation}
\phi_N\sim \exp(-W_0(x))(z-a_1)^{p_1}(z-a_2)^{p_2}\prod_{k=1}^N
(z-z_k)
 \end{equation}

Now Eqs.~(\ref{V0p}) and (\ref{dVp}) are generalized to
\begin{eqnarray}
V_0&=&W_0^{\prime 2}-W_0^{\prime\prime}-2\left(W_0^\prime z^\prime
-\frac{z^{\prime\prime}}{2}\right)\left(\frac{p_1}{z-a_1}+\frac{p_2}{z-a_2}\right)
\nonumber\\
&&+ z^{\prime 2}\left[\frac{p_1(p_1-1)}{(z-a_1)^2}+
\frac{p_2(p_2-1)}{(z-a_2)^2}+\frac{2p_1p_2}{(z-a_1)(z-a_2)}\right],
\label{V0-2s}\\
 \Delta V_N &=& -2\left(W_0^\prime z^\prime
-\frac{z^{\prime\prime}}{2}\right)\sum_{k=1}^N
\frac{1}{z-z_k}\nonumber\\
&& + z^{\prime 2}\left[\sum_{k,l\atop k\neq l}
\frac{1}{(z-z_k)(z-z_l)} + 2\left(\frac{p_1}{z-a_1}+
\frac{p_2}{z-a_2}\right)\sum_{k=1}^N \frac{1}{z-z_k} \right].
\label{dV-2s}
\end{eqnarray}
Once again, the system is completely defined by the choice of
$z^{\prime 2}$ and $W_0$.  The analysis of the solvability of the
system is more complicated than before, but the principle is the
same.   For polynomial $z^{\prime 2}$ and $W_0^\prime z^\prime$
the systems are new QES models in general.

As before let us take $z^{\prime 2}$ to be at most quadratic in
$z$.  A situation of interest is that in which $a_1$ and $a_2$ are
the two real zeros of $Q_2(z)$, i.e., if $z^{\prime 2}= A
(z-a_1)(z-z_2)$ where $A$ is real.  In this case the sufficient
conditions for ES and QES models are the same as before, namely,
one could get ES system if $m\leq 1$, and QES otherwise.  But just
as the Morse potential discussed in the last section, the QES
system could become an ES one if $p_1$ and $p_2$ take on special
values. The ES Rosen-Morse I and II potentials and the Eckart
potential in \cite{Cooper} are such cases.

We shall illustrate the construction of a QES model first
discussed in \cite{ST} (see also \cite{HS}). This model was not
listed in \cite{Tur} but is found to be also related to $sl(2)$
algebra \cite{HS}.  The function $z(x)$ is taken to be
$z(x)=\sin^2 x$. This is a solution of $z^{\prime 2}=4z(1-z)$, and
hence $a_1=0$ and $a_2=1$, and $0<x<\pi/2$.  In order to obtain a
type 1 QES model we should choose $W_0$ such that $W_0^\prime
z^\prime$ is of the second degree in $z$.  Let us take
$P_2(z)=4az(z-1)$, where $a$ is real. This gives a solution
$W_0(x)=a\cos 2x/2$.  With the chosen $Q_2(z)$ and $P_2(z)$, we
obtain from Eq.~(\ref{V0-2s}) and (\ref{dV-2s}) the following QES
potential
\begin{eqnarray}
V_N(x)&=&a^2 \sin^2 2x + 2a\cos 2x + 2\left(a\sin^2 2x+\cos
2x\right)\left(\frac{p_1}{\sin^2
x}-\frac{p_2}{\cos^2 x}\right)\nonumber\\
&& +4p_1(p_1-1)\cot^2 x + 4p_2(p_2-1)\tan^2 x\\
&&-8aN\sin^2 x -8a\sum_{k=1}^N z_k
-4N\left[N-2(a-p_1-p_2)\right]-8p_1 p_2,\nonumber
\end{eqnarray}
where the $z_k$'s satisfy the BAE ($k=1,2,\ldots,N$):
\begin{equation}
4az_k^2 -2\left(2(a-p_1-p_2)-1\right)z_k -1 -
4z_k(1-z_k)\sum_{l\neq k} \frac{1}{z_k-z_l}=0.
\end{equation}
The forms of $V_N$ and BAE in \cite{HS} for this model is regained
by identifying $p_1=c/2-b/4$ and $p_2=b/4$, where $b$ and $c$ are
the parameters used in that paper.


\section{Summary}

We have discussed exact and quasi-exact solvability of the
schr\"odinger equation based on the approach of prepotential.
Three types of $N$-th order prepotentials are described which
cover most of the known ES and QES models, and which are capable
of generating new ones.  It is shown that the choice of coordinate
transformation $z^{\prime 2}(x)$ and the zero-th order
prepotential $W_0(x)$ completely determine the form of $V_N$.
General conditions on the choice of $z^{\prime 2}$ and $W_0(x)$
for exact and quasi-exact solvabilities were described. These
conditions were illustrated by various examples.  The prepotential
approach is quite elegant in itself. Moreover, it allows easy
extensions of ES and QES theory to systems with multi-component
wave functions, such as the Pauli and Dirac equations, and to the
Fokker-Planck equations, as prescribed in
\cite{HoRoy1,Ho4,Ho5,HS}.

In this work we have confined our discussions to ES and QES models
involving a change of coordinates to the so-called sinusoidal
coordinates. These are coordinates $z(x)$ which satisfy $z^{\prime
2}=Q_2(z)$, or $z^{\prime\prime}=R_0(z)$ or $R_1(z)$. These
coordinates cover most of the known ES and QES models. The
examples we presented here by no means exhaust all possible
sinusoidal coordinates.  Other QES cases admitted by certain
sinusoidal transformations need be studied.

But as discussed in Sect.~II, to construct a type 1 QES system it
is also possible to choose $z^{\prime 2}$ to be a cubic
polynomial, i.e. $z^{\prime 2}=Q_3(z)$ . This lies beyond the
transformations by sinusoidal coordinates. Such transformations
are less studied in the literature. However, they could give rise
to new QES systems, and deserve further study. In fact,  two
$sl(2)$-based QES cases in \cite{Tur}, namely, class IV and V,
require $n=3$ and $m=1$ and $3$, respectively.

Only three types of $N$-th order prepotentials were discussed
here, namely, prepotentials for systems defined on the whole line,
on the half-line, and on a finite interval. Generalizing to
several singularities is straightforward.

We have only considered cases for which $W_0^\prime z^\prime$ and
$z^{\prime 2}$ are polynomials in $z$. It is interesting to
consider other possibilities for these two defining functions.

Finally, it is also interesting to extend the present approach to
systems with non-Hermitian Hamiltonians admitting real spectra. A
preliminary attempt was presented in \cite{Ho-CNY85}, where some
non-Hermitian QES Hamiltonians, including that given in
\cite{BB1}, were generated by making some coefficients in $P_m(z)$
complex.  However, a full treatment of this kind of systems
usually requires one to extend the basic variable $x$ to the
complex plane \cite{BB2,B}. Hence a better understanding of the
prepotential approach to non-Hermitian Hamiltonians is needed.

\begin{acknowledgments}

I thank R. Sasaki for stimulating discussions and for bringing my
attention to sinusoidal coordinates. This work is supported in
part by the National Science Council (NSC) of the Republic of
China under Grant Nos. NSC 96-2112-M-032-007-MY3 and NSC
95-2911-M-032-001-MY2.

\end{acknowledgments}



\begin{thebibliography}{99}

\bibitem{TU}  A.V. Turbiner and A.G. Ushveridze, Phys. Lett. {\bf A126}
(1987) 181.

\bibitem{Tur1} A.G. Turbiner, Sov. Phys. JETP {\bf 67} (1988) 230.

\bibitem{Tur} A.V. Turbiner, Comm. Math. Phys. {\bf 118} (1988) 467.

\bibitem{Ush1} A.G. Ushveridze, {\sl Sov. Phys.-Lebedev Inst. Rep.} {\bf
2} (1988) 50; 54.

\bibitem{Ush} A.G. Ushveridze, {\sl Quasi-exactly solvable models in quantum
mechanics} (IOP, Bristol, 1994).

\bibitem{O1} N. Kamran and P.J. Olver, J. Math. Anal. Appl. {\bf 145} (1990)
342.

\bibitem{O2} A. Gonz\'alez-L\'opez, N. Kamran and P.J. Olver, Comm. Math. Phys.
{\bf 153} (1993) 117.

\bibitem{O3} A. Gonz\'alez-L\'opez, N. Kamran and P.J. Olver, Contemp. Math.  {\bf 160} (1994) 113.

\bibitem{Tur2}  M.A. Shifman and A.V. Turbiner, Comm. Math. Phys. {\bf
126} (1989) 347.

\bibitem{Tur3} A.V. Turbiner, Contemp. Math. {\bf 160} (1994) 263.

\bibitem{Tur4} G. Post and A.V. Turbiner, Russ. J. Math. Phys. {\bf 3} (1995)
113.

\bibitem{S1} M.A. Shifman, Int. J. Mod. Phys. {\bf A4} (1989) 2897; 3305.

\bibitem{S2} M.A. Shifman, Contemp. Math. {\bf 160} (1994) 237.

\bibitem{HoRoy1} C.L. Ho and P. Roy, J. Phys. {\bf A36} (2003) 4617.

\bibitem{B1} Y. Brihaye and P. Kosinski, Mod. Phys. Lett. {\bf A13} (1998)
1445.

\bibitem{B2} Y. Brihaye and A. Nininahazwe, Mod. Phys. Lett. {\bf
A20} (2005) 1875.

\bibitem{Znojil} M. Znojil, Mod. Phys. lett. {\bf A14} (1999) 863.

\bibitem{Ho1}
C.-L. Ho and V.R. Khalilov, Phys. Rev. {\bf A61} (2000) 032104.

\bibitem{Ho2} C.-M. Chiang and C.-L. Ho, J. Math. Phys. {\bf 43} (2002)
43.

\bibitem{Ho3} C.-M. Chiang and C.-L. Ho, Mod. Phys. Lett. {\bf A20} (2005)
673.

\bibitem{Ho4} C.-L. Ho and P. Roy, Ann. Phys. {\bf 312} (2004)
161.

\bibitem{Ho5} C.-L. Ho, Ann. Phys. {\bf 321} (2006) 2170.

\bibitem{SH} A. Schulze-Halberg, Chinese J. Phys. {\bf 42} (2004)
117; 234.

\bibitem{CH}
H.-T. Cho and C.-L. Ho, J. Phys {\bf A40} (2007) 1325.

\bibitem{Gaudin}
For a first systematic discussion of the use of Bethe ansatz
equations, see e.g., M. Gaudin, {\sl La Fonction d'Onde de Bethe}
(Masson, Paris, 1983).

\bibitem{PT} G. Post and A. Turbiner, Russian J. Math. Phys. {\bf
3} (1995) 113.

\bibitem{KMO} N. Kamran, R. Milson and P.J.
Olver, Invariant modules and the reduction of nonlinear partial
differential equations to dynamical systems (1999).
solv-int/9904014.

\bibitem{ST} R. Sasaki and K. Takasaki, J. Phys. {\bf A34} (2001)
9533.

\bibitem{Cooper}  F. Cooper, A. Khare and U. Sukhatme, Phys. Rep. {\bf 251}
(1995) 267.

\bibitem{Junker} G. Junker, {\sl Supersymmetric Methods in Quantum
and Statistical Physics} (Springer-Verlag, Berlin, 1996).

\bibitem{CS} E. Corrigan and R. Sasaki, J.Phys. {\bf A35} (2002)
7017.

\bibitem{Sa1}
A.J. Bordner, N.S. Manton and R. Sasaki, Prog. Theor. Phys. {\bf
103} (2000) 463.

\bibitem{Sa2} S.P. Khastgir, A.J. Pocklington and R. Sasaki,
 J. Phys. {\bf A33} (2000) 9033.

\bibitem{Sa3} S. Odake and R. Sasaki, J. Phys. {\bf A35} (2002) 8283.

\bibitem{Sa4}I. Loris and R. Rasaki, J. Phys. {\bf A37} (2004) 211.

\bibitem{Sa5} R. Sasaki and K. Takasaki, J. Math. Phys. {\bf 47} (2006) 012701.

\bibitem{HS} C.-L. Ho and R. Sasaki, Qausi-exactly solvable
Fokker-Planck equations, Tamkang preprint and Yukawa Institute
(Kyoto Univ.) Report No.Y ITP-07-21 (to appear in Ann. Phys.)
arXiv:0705.0863.v2.

\bibitem{fac1}
E. Schr\"odinger, Proc. Roy. Irish Acad. {\bf A46} (1940) 9.

\bibitem{fac2}
L.Infeld and T.E. Hull, Rev. Mod. Phys. {\bf 23} (1951) 21.

\bibitem{OS} S. Odake and R. Sasaki, J. Math. Phys. {\bf 47} (2006)
102102.

\bibitem{Stiel}
T.J. Stieltjies, ``Sur quelques th\'eor\`emes d'alg\`ebre", Compte
Rendus {\bf 100} (1885) 439-440.  In Stieltjies' aproach, the
zeros of classical polynomials can be interpreted as the
equilibrium positions of a related electrostatic system.  Such
analogy was rediscovered in \cite{Ush1}.

\bibitem{Szego}
G.~Szeg\"o, {\sl Orthogonal Polynomials}, Amer. Math. Soc.
Colloquium Publications Vol. 23 (Amer. Math. Soc., New York,
1939).

\bibitem{Ho-CNY85}
C.-L. Ho, Prepotential approach to exact and quasi-exact
solvabilities of Hermitian and non-Hermitian Hamiltonians. Talk
presented at ``Conference in Honor of CN Yang's 85th Birthday", 31
Oct - 3 Nov, 2007, Singapore. arXiv:0801.0944 [hep-th].

\bibitem{BB1} C.M. Bender and S. Boettcher, J. Phys. {\bf A31} (1998) L273.

\bibitem{BB2} C.M. Bender and S. Boettcher, Phys. Rev. Lett. {\bf
80} (1998) 5234.

\bibitem{B} C.M. Bender, Contemp.Phys. {\bf 46} (2005) 277.

\end{thebibliography}
\end{document}